\documentclass{PoS}

\title{Pion-assisted $N\Delta$ and $\Delta\Delta$ dibaryons\thanks{Work 
supported in part by the EU initiative FP7, HadronPhysics3, under the SPHERE 
and LEANNIS cooperation programs, and in part by COFAA-IPN (M\'exico).}} 

\ShortTitle{Pion-assisted dibaryons} 

\author{\speaker{Avraham Gal} \\ Racah Institute of Physics, 
The Hebrew University, 91904 Jerusalem, Israel \\ 
E-mail: \email{avragal@savion.huji.ac.il}}

\author{Humberto Garcilazo \\ Escuela Superior de F\' \i sica y Matem\'aticas, 
Instituto Polit\'ecnico Nacional, Edificio 9, 07738 M\'exico D.F., Mexico \\ 
E-mail: \email{humberto@esfm.ipn.mx}} 

\abstract{$N\Delta$ and $\Delta\Delta$ dibaryon candidates are discussed and 
related quark-based calculations are reviewed. New hadronic calculations of 
$L=0$ nonstrange dibaryon candidates are reported. For $N\Delta$, $I(J^P)=
1(2^+)$ and $2(1^+)$ $S$-matrix poles slightly below threshold are found by 
solving $\pi NN$ Faddeev equations with relativistic kinematics, and for 
$\Delta\Delta$ several $S$-matrix poles below threshold are found by solving 
$\pi N\Delta$ Faddeev equations with relativistic kinematics in which the 
$N\Delta$ interaction is dominated by the $1(2^+)$ and $2(1^+)$ resonating 
channels. In particular, the $I(J^P)=0(3^+)$ $\Delta\Delta$ dibaryon candidate 
${\cal D}_{03}(2370)$ observed recently by the WASA@COSY Collaboration 
is naturally explained in terms of long-range physics dominated by pions, 
nucleons and $\Delta$'s. These calculations are so far the only ones to 
reproduce the relatively small width $\approx$70~MeV of ${\cal D}_{03}(2370)$. 
Predictions are also made for the location and width of ${\cal D}_{30}$, the 
$I(J^P)=3(0^+)$ exotic partner of ${\cal D}_{03}(2370)$.}

\FullConference{XV International Conference on Hadron Spectroscopy-Hadron 
2013\\ 4-8 November 2013\\ Nara, Japan}

\begin{document}

\section{Introduction}

QCD-motivated studies of six-quark ($6q$) dibaryons started with Jaffe's 
prediction of the deeply bound $uuddss$ $I(J^P)=0(0^+)$ $H$ dibaryon 
\cite{jaffe77} using the color-magnetic (CM) one-gluon exchange interaction 
$V_{CM}=\sum_{i<j}-(\lambda_i\cdot\lambda_j)(s_i\cdot s_j)v(r_{ij})$, where 
$v(r_{ij})$ is a flavor conserving $qq$ short-range potential which for a 
totally symmetric $L=0$ wavefunction is approximated by its matrix element 
${\cal M}_0$. From the $\Delta$ -- $N$ mass difference of $\approx$300~MeV, 
one estimates ${\cal M}_0\sim$75~MeV. Leading dibaryon candidates for 
strangeness $\cal S$ ranging from 0 to $-$3 are listed in Table~\ref{tab:oka}, 
where $\delta <V_{CM}>$ is the contribution of $V_{CM}$ to the dibaryon mass 
with respect to its contribution to the dibaryon's constituents $B$ and $B'$. 
Dibaryon candidates with quarks heavier than $s$ are not covered here. 

\begin{table}[hbt]  
\begin{center} 
\begin{tabular}{lcclcc} 
\hline 
$\cal S$ & SU(3)$_{\rm f}$ & $I$ & $J^{\pi}$ & $BB'$ structure & 
$\delta <V_{CM}>$  \\ \hline 
~0 & [3,3,0] $\overline {\bf 10}$ & 0 & $3^+$ & $\Delta\Delta$ & 0 \\ 
$-$1 & [3,2,1] ${\bf 8}$ & 1/2 & $2^+$ & 
${\sqrt {1/5}}~(N\Sigma^*+2~\Delta\Sigma)$ & $-{\cal M}_0$ \\ 
$-$2 & [2,2,2] ${\bf 1}$ & 0 & $0^+$ & ${\sqrt {1/8}}~(\Lambda\Lambda+2~N\Xi-
{\sqrt 3}~\Sigma\Sigma)$ & $-2{\cal M}_0$ \\ 
$-$3 & [3,2,1] ${\bf 8}$ & 1/2 & $2^+$ & ${\sqrt {1/5}}~[{\sqrt 2}~N\Omega-
(\Lambda\Xi^*-\Sigma^*\Xi+\Sigma\Xi^*)]$ & $-{\cal M}_0$ \\ 
\hline 
\end{tabular} 
\caption{Leading quark-based $L=0$ dibaryon candidates, adapted from 
Ref.~\cite{gal11}.} 
\label{tab:oka} 
\end{center} 
\end{table} 

Let's comment on the two extreme cases in the table: ${\cal S}=-2$ and 
${\cal S}=0$. For ${\cal S}=-2$, the table suggests that the listed $H$ 
dibaryon is deeply bound, located well below the $\Lambda\Lambda$ lowest 
particle-stability threshold, but in fact SU(3)$_{\rm f}$ breaking effects 
abort its anticipated stability, as concluded recently from chiral 
extrapolations of lattice QCD calculations \cite{lattice13}; see also 
\cite{gal13} for arguments based on hypernuclear phenomenology. For 
${\cal S}=0$, in contrast, the table suggests no outstanding nonstrange 
dibaryon candidate resulting from the quark-based CM interaction.
However, $N\Delta$ and $\Delta\Delta$ $s$-wave dibaryon resonances 
${\cal D}_{IS}$ with isospin $I$ and spin $S$ were proposed as early as 
1964, when quarks were still perceived as merely mathematical entities, 
by Dyson and Xuong \cite{dyson64} who focused on the lowest-dimension 
SU(6) multiplet in the $\bf{56\times 56}$ product that contains the SU(3) 
$\overline{\bf 10}$ and ${\bf 27}$ multiplets in which the deuteron 
${\cal D}_{01}$ and $NN$ virtual state ${\cal D}_{10}$ are classified. 
This yields two dibaryon candidates, ${\cal D}_{12}$ for $N\Delta$ and 
${\cal D}_{03}$ for $\Delta\Delta$ listed in Table~\ref{tab:dyson} with 
masses $M=A+B[I(I+1)+S(S+1)-2]$ in terms of constants $A,B$. Identifying 
$A$ with the $NN$ threshold mass 1878~MeV, $B\approx 47$~MeV was determined 
by assigning ${\cal D}_{12}$ to the $pp\leftrightarrow \pi^+ d$ reaction 
channels resonating at 2160~MeV near the $N\Delta$ threshold. This led to 
a predicted mass value of 2350~MeV for ${\cal D}_{03}$. The ${\cal D}_{03}$ 
dibaryon was the subject of many quark-based model calculations since 1980. 

\begin{table}[hbt]
\begin{center}
\begin{tabular}{ccccccc}
\hline
${\cal D}_{IS}$ & ${\cal D}_{01}$ & ${\cal D}_{10}$ & ${\cal D}_{12}$ &
${\cal D}_{21}$ & ${\cal D}_{03}$ & ${\cal D}_{30}$ \\
\hline
$BB'$ &$NN$&$NN$& $N\Delta$ & $N\Delta$ & $\Delta\Delta$ & $\Delta\Delta$ \\
SU(3)$_{\rm f}$ & $\overline{\bf 10}$ & ${\bf 27}$ & ${\bf 27}$ & ${\bf 35}$ &
$\overline{\bf 10}$ & ${\bf 28}$ \\
$M({\cal D}_{IS})$ & $A$ & $A$ & $A+6B$ & $A+6B$ & $A+10B$ & $A+10B$ \\
\hline
\end{tabular}
\caption{SU(6) predictions \cite{dyson64} for nonstrange $L=0$ dibaryons
${\cal D}_{IS}$ with isospin $I$ and spin $S$.}
\label{tab:dyson}
\end{center} 
\end{table} 

\begin{table}[hbt] 
\begin{center} 
\begin{tabular}{cccccccccc}
\hline
$M$(GeV) & \cite{dyson64} & \cite{muld80} & \cite{oka80} & \cite{muld83} 
& \cite{malt85} & \cite{gold89} & \cite{sal02} & \cite{QDCSM09,QDCSM13} & 
exp. \\ 
\hline 
${\cal D}_{03}$ ($\Delta\Delta$) & 2.35 & 2.36 & 2.46 & 2.38 & 2.20 & 
$\leq$2.26 & 2.46 & 2.36$^\ast$ & 2.37 \cite{wasa08,wasa11} \\ 
${\cal D}_{12}$ ($N\Delta$) & 2.16$^{\ast\ast}$ & 2.36 & -- & 2.36 & -- & -- & 
2.17 & -- & $\approx$2.15 \cite{arndt87,hosh92} \\  
\hline
\end{tabular}
\caption{Quark-based model predictions of ${\cal D}_{03}$ and ${\cal D}_{12}$, 
except where denoted by asterisks: $^\ast$ denotes post-experiment 
${\cal D}_{03}$ calculation and $^{\ast\ast}$ denotes input from experiment. 
Experimental evidence for ${\cal D}_{03}(2370)$ is shown in the figure below. 
The $N\Delta$ and $\Delta\Delta$ thresholds are at 2.17 and 2.46 GeV, 
respectively.} 
\label{tab:D03D12calcs} 
\end{center} 
\end{table} 

\begin{figure}[htb] 
\begin{center} 
\includegraphics[width=0.48\textwidth,height=4.5cm]{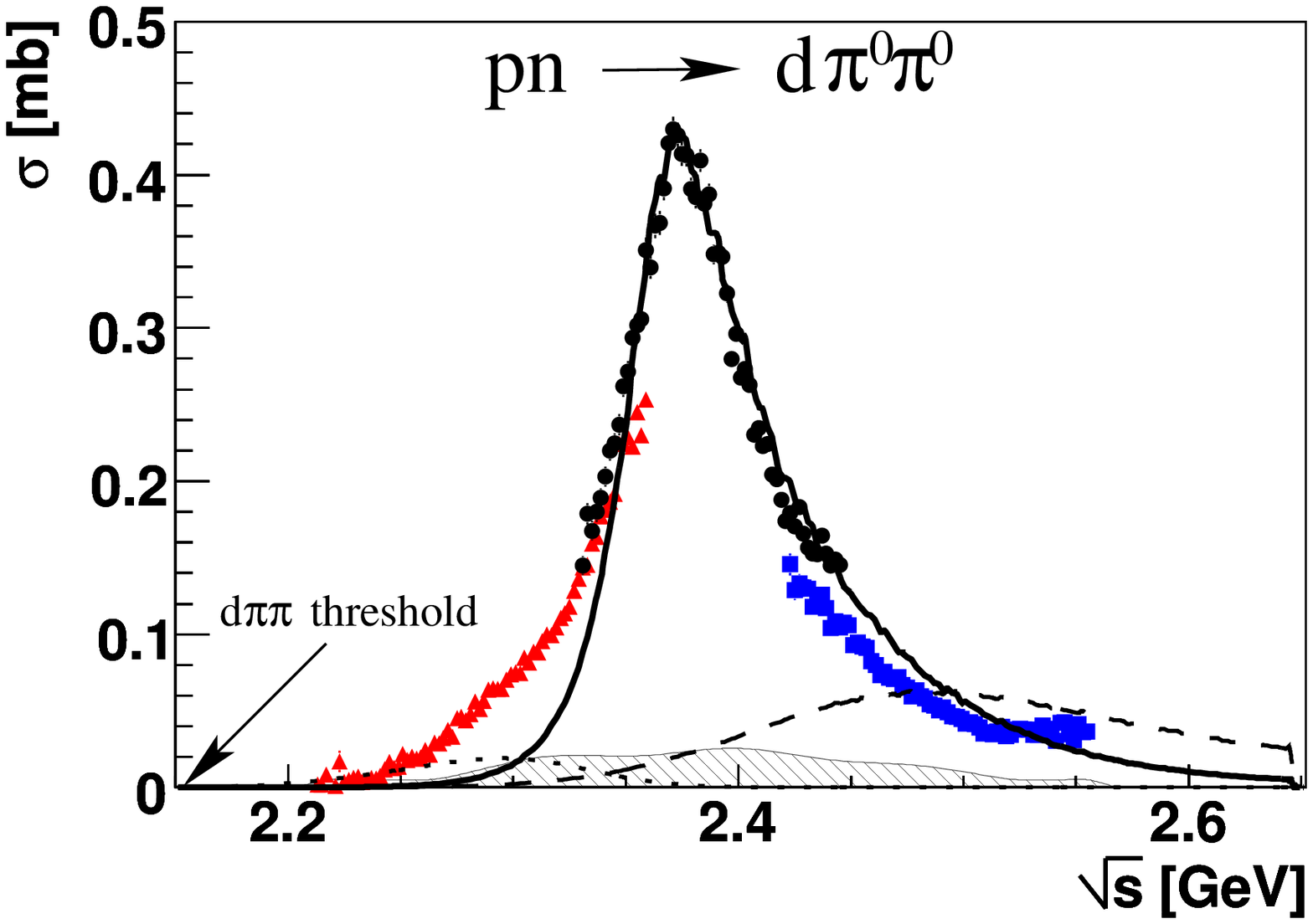} 
\includegraphics[width=0.48\textwidth,height=4.5cm]{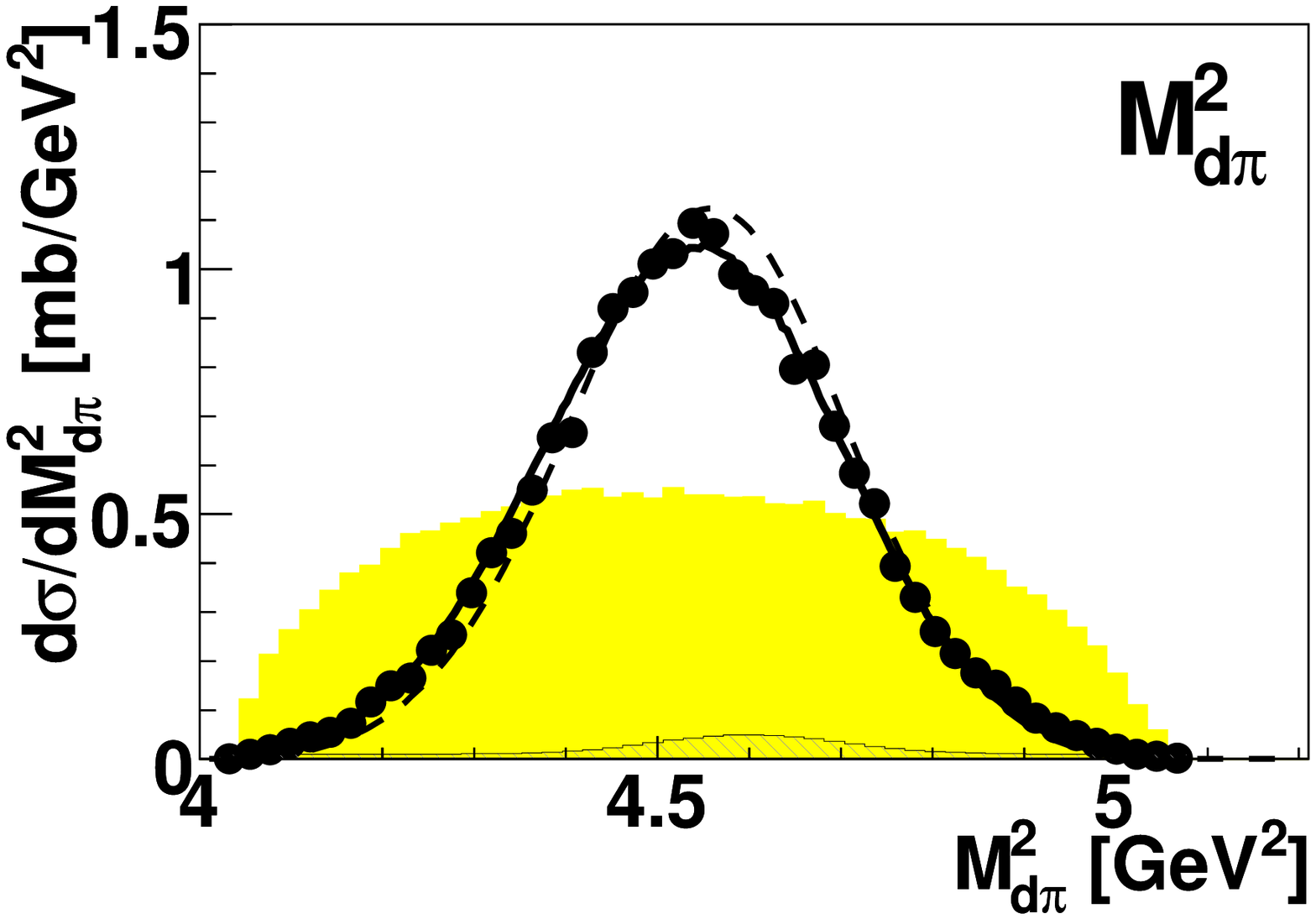} 
\caption{${\cal D}_{03}(2370)$ $\Delta\Delta$ dibaryon signal on the left 
panel, and its $M^2_{d\pi^0}$ Dalitz-plot projection on the right panel, from 
$pn\to d\pi^0\pi^0$ by WASA-at-COSY \cite{wasa11}. Figures courtesy of Heinz 
Clement.} 
\label{fig:data} 
\end{center} 
\end{figure}

${\cal D}_{03}$ mass predictions are listed in Table~\ref{tab:D03D12calcs} 
for several representative approaches. The table exhibits a broad range of 
caluclated ${\cal D}_{03}$ masses. Except for the Dyson-Xuong pioneering 
prediction \cite{dyson64} none of those confronting ${\cal D}_{03}$ and 
${\cal D}_{12}$ succeeded to correctly reproduce both. Recent experimental 
evidence for ${\cal D}_{03}$ is displayed in Fig.~\ref{fig:data}--left. 
Isospin $I=0$ is uniquely fixed in this two-pion production reaction and 
a spin-parity $3^+$ assignment follows from the measured deuteron and pions 
angular distributions, assuming $s$-wave decaying $\Delta\Delta$ pair. 
The peak of the $M^2_{d\pi^0}$ distribution on the right panel at 
$\sqrt{s}\approx$2.13~GeV, almost at the ${\cal D}_{12}$ $N\Delta$ dibaryon 
peak, suggests that ${\cal D}_{12}$ plays a role in forming the $\Delta\Delta$ 
dibaryon ${\cal D}_{03}$. It is shown below that the pion-assisted methodology 
applied by us recently \cite{gg13,gg14} couples the two dibaryons dynamically 
in a more natural way than appears in quark-based models. Our calculations 
emphasize the long-range physics aspects of nonstrange dibaryons, 
as described briefly in the next section. 

\section{Pion-assisted nonstrange dibaryons} 
\subsection{$N\Delta$ dibaryons} 

\begin{figure}[hbt] 
\begin{center} 
\includegraphics[width=0.6\textwidth]{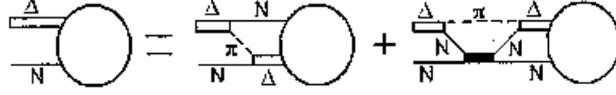} 
\caption{$N\Delta$ dibaryon's Lippmann-Schwinger equation \cite{gg14}.} 
\label{fig:DIS} 
\end{center} 
\end{figure}

The ${\cal D}_{12}$ dibaryon shows up experimentally as $NN({^1D_2})$ 
$\leftrightarrow$ $\pi d({^3P_2})$ coupled-channel resonance corresponding 
to a quasibound $N\Delta$ with mass $M\approx 2.15$~GeV, near the $N\Delta$ 
threshold, and width $\Gamma\approx 0.12$~GeV \cite{arndt87,hosh92}. In our 
recent work \cite{gg14} we have calculated this dibaryon and other $N\Delta$ 
dibaryon candidates such as ${\cal D}_{21}$ (see Table~\ref{tab:dyson}) 
by solving Faddeev equations with relativistic kinematics for the $\pi NN$ 
three-body system, where the $\pi N$ subsystem is dominated by the $P_{33}$ 
$\Delta$(1232) resonance channel and the $NN$ subsystem is dominated by 
the $^3S_1$ and $^1S_0$ channels. The coupled Faddeev equations give rise 
then to an effective $N\Delta$ Lippmann-Schwinger (LS) equation for the 
three-body $S$-matrix pole, with energy-dependent kernels that incorporate 
spectator-hadron propagators, as shown diagrammatically in Fig.~\ref{fig:DIS} 
where circles denote the $N\Delta$ $T$ matrix. 

Of the four possible $L=0$ $N\Delta$ dibaryon candidates ${\cal D}_{IS}$ 
with $IS=12,21,11,22$, the latter two do not provide resonant solutions. 
For ${\cal D}_{12}$, only $^3S_1$ contributes out of the two $NN$ 
interactions, while for ${\cal D}_{21}$ only $^1S_0$ contributes. 
Since the $^3S_1$ interaction is the more attractive one, ${\cal D}_{12}$ 
lies below ${\cal D}_{21}$ as borne out by the calculated masses listed 
in Table~\ref{tab:NDel} for two choices of the $P_{33}$ interaction form 
factor corresponding to spatial sizes of 1.35~fm and 0.9~fm of the $\Delta$ 
isobar. The two dibaryons are found to be degenerate to within less than 
20~MeV. The mass values calculated for ${\cal D}_{12}$ are reasonably close 
to the value $W=2148-{\rm i}63$~MeV \cite{arndt87} and $W=2144-{\rm i}55$~MeV 
\cite{hosh92} derived in coupled-channel phenomenological analyses. 

\begin{table}[hbt] 
\begin{center} 
\begin{tabular}{ccccc} 
\hline 
$W^{>}({\cal D}_{12})$ & $W^{>}({\cal D}_{21})$ &  &
$W^{<}({\cal D}_{12})$ & $W^{<}({\cal D}_{21})$   \\  
\hline 
2147$-{\rm i}$60 & 2165$-{\rm i}$64 &  &  2159$-{\rm i}$70 & 
2169$-{\rm i}$69 \\
\hline 
\end{tabular} 
\caption{$N\Delta$ dibaryon $S$-matrix poles (in MeV) for ${\cal D}_{12}$ 
and ${\cal D}_{21}$, obtained by solving $\pi NN$ Faddeev equations for two 
choices of the $\pi N$ $P_{33}$ form factor, with large (small) spatial size 
denoted > (<).} 
\label{tab:NDel} 
\end{center} 
\end{table} 

\subsection{$\Delta\Delta$ dibaryons} 

\begin{figure}[htb] 
\begin{center} 
\includegraphics[width=0.6\textwidth]{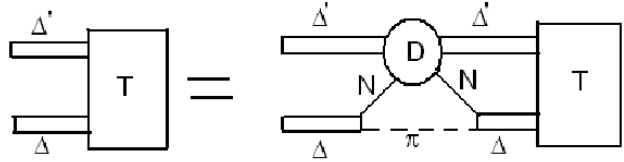} 
\caption{$S$-matrix pole equation for ${\cal D}_{03}(2370)$ $\Delta\Delta$ 
dibaryon \cite{gg13}.} 
\label{fig:D03} 
\end{center} 
\end{figure} 

Four-body $\pi\pi NN$ calculations are required, strictly speaking, 
to discuss $\Delta\Delta$ dibaryons. In Ref.~\cite{gg13} we studied the 
${\cal D}_{03}$ dibaryon by solving a $\pi N\Delta'$ three-body model, 
where $\Delta'$ is a stable $\Delta$(1232) and the $N\Delta'$ interaction 
is dominated by the ${\cal D}_{12}$ dibaryon. The $I(J^P)=1(2^+)$ 
$N\Delta'$ interaction was not assumed to resonate but, rather, it was 
fitted within a $NN$--$\pi NN$--$N\Delta'$ coupled-channel caricature 
model to the $NN$ $^1D_2$ $T$-matrix, requiring that the resulting 
$N\Delta'$ separable-interaction form factor is representative of long-range 
physics, with momentum-space soft cutoff $\Lambda\lesssim$3~fm$^{-1}$. 
The $\pi N$ interaction was again assumed to be dominated by the $P_{33}$ 
$\Delta$ resonance. The Faddeev equations of this three-body model give rise, 
as before, to an effective LS equation for the $\Delta\Delta'$ $S$-matrix pole 
corresponding to ${\cal D}_{03}$. This LS equation is shown diagrammatically 
in Fig.~\ref{fig:D03}, where $D$ stands for the ${\cal D}_{12}$ dibaryon. 
In Ref.~\cite{gg14} we have extended the calculation of ${\cal D}_{03}$ 
to other ${\cal D}_{IS}$ $\Delta\Delta$ dibaryon candidates, with $D$ now 
standing for both $N\Delta$ dibaryons ${\cal D}_{12}$ and ${\cal D}_{21}$. 
Since ${\cal D}_{21}$ is almost degenerate with ${\cal D}_{12}$, and 
with no $NN$ observables to constrain the input $(I,S)$=(2,1) $N\Delta'$ 
interaction, the latter was taken the same as for $(I,S)$=(1,2). The model 
dependence of this assumption is under study at present. The lowest and 
also narrowest $\Delta\Delta$ dibaryons found are ${\cal D}_{03}$ and 
${\cal D}_{30}$.

\begin{table}[hbt] 
\begin{center} 
\begin{tabular}{ccccccc} 
\hline 
$W(\Delta')$ & $W^{>}({\cal D}_{03})$ & $W^{>}({\cal D}_{30})$ & 
$W^{<}({\cal D}_{03})$ & $W^{<}({\cal D}_{30})$ & $W_{\rm av}({\cal D}_{03})$ 
& $W_{\rm av}({\cal D}_{30})$  \\  
\hline 
1211$-{\rm i}$49.5 & 2383$-{\rm i}$47 & 2412$-{\rm i}$49 & 
2342$-{\rm i}$31 & 2370$-{\rm i}$30 & 2363$-{\rm i}$39 & 2391$-{\rm i}$39 \\
1211$-{\rm i}$(2/3)49.5 & 2383$-{\rm i}$41 & 2411$-{\rm i}$41 & 
2343$-{\rm i}$24 & 2370$-{\rm i}$22 & 2363$-{\rm i}$33 & 2390$-{\rm i}$32 \\
\hline 
\end{tabular} 
\caption{$\Delta\Delta$ dibaryon $S$-matrix poles (in MeV) obtained in 
Refs.~[18,19] by using a spectator-$\Delta'$ complex mass $W(\Delta')$ 
(first column) in the propagator of the LS equation depicted in Fig.~3. 
The last two columns give calculated mass and width values averaged over 
those from the > and < columns, where > and < are defined in the caption 
of Table~4.} 
\label{tab:DelDel} 
\end{center} 
\end{table} 

Representative results for ${\cal D}_{03}$ and ${\cal D}_{30}$ are assembled 
in Table~\ref{tab:DelDel}, where the calculated mass and width values listed 
in each row correspond to the value listed there of the spectator-$\Delta'$ 
complex mass $W(\Delta')$ used in the propagator of the LS equation shown in 
Fig.~\ref{fig:D03}. The value of $W(\Delta')$ in the first row is that of the 
$\Delta$(1232) $S$-matrix pole. It is implicitly assumed thereby that the 
decay $\Delta' \to N\pi$ proceeds independently of the $\Delta \to N\pi$ 
isobar decay. However, as pointed out in Ref.~\cite{gg13}, care must be 
exercised to ensure that the decay nucleons and pions satisfy Fermi-Dirac 
and Bose-Einstein statistics requirements, respectively. Assuming $L=0$ for 
the decay-nucleon pair, this leads to the suppression factor 2/3 depicted 
in the value of $W(\Delta')$ listed in the second row. It is seen that the 
widths obtained upon applying this width-suppression are only moderately 
smaller, by less than 15 MeV, than those calculating disregarding this 
quantum-statistics correlation. 

The mass and width values calculated for ${\cal D}_{03}$ \cite{gg13} 
agree very well with those determined by the WASA-at-COSY Collaboration 
\cite{wasa11}, reproducing in particular the reported width value 
$\Gamma({\cal D}_{03})=68$~MeV which is extremely low with respect to the 
expectation $\Gamma_{\Delta}\leq\Gamma({\cal D}_{03})\leq 2\Gamma_{\Delta}$, 
with $\Gamma_{\Delta}\approx 118$~MeV. No other calculation has succeeded 
so far to do that. Similar small widths according to Table~\ref{tab:DelDel} 
hold for ${\cal D}_{30}$ which is located about 30 MeV above ${\cal D}_{03}$. 
This is about half of the spacing found very recently in the quark-based 
calculations of Ref.~\cite{QDCSM13}. Note, however, that the widths calculated 
there are considerably larger than ours. A more complete discussion of these 
and of other ${\cal D}_{IS}$ $\Delta\Delta$ dibaryon candidates is found in 
Ref.~\cite{gg14}.

\section{Conclusion} 

It was shown how the 1964 Dyson-Xuong SU(6)-based classification and 
predictions of nonstrange dibaryons \cite{dyson64} are confirmed in our 
hadronic model of pion-assisted $N\Delta$ and $\Delta\Delta$ dibaryons 
\cite{gg13,gg14}. The input for dibaryon calculations in this model consists 
of nucleons, pions and $\Delta$'s, interacting via long-range pairwise 
interactions. These calculations reproduce the two nonstrange dibaryons 
established experimentally and phenomenologically so far, the $N\Delta$ 
dibaryon ${\cal D}_{12}$ \cite{arndt87,hosh92} and the $\Delta\Delta$ dibaryon 
${\cal D}_{03}$ reported by the WASA-at-COSY Collaboration \cite{wasa11}, 
and also predict an exotic $I=2$ $N\Delta$ dibaryon ${\cal D}_{21}$ nearly 
degenerate with ${\cal D}_{12}$. We note that ${\cal D}_{12}$ provides in our 
$\pi N\Delta$ three-body model of ${\cal D}_{03}$ a two-body decay channel 
$\pi {\cal D}_{12}$ with threshold lower than $\Delta\Delta$. Our calculations 
are capable of dealing with other $\Delta\Delta$ dibaryon candidates, 
in particular the $I=3$ exotic ${\cal D}_{30}$ highlighted recently by 
Bashkanov, Brodsky and Clement \cite{BBC13}. These authors emphasized the 
dominant role that 6q hidden-color configurations might play in binding 
${\cal D}_{03}$ and ${\cal D}_{30}$, but recent explicit quark-based 
calculations \cite{QDCSM13} find these configurations to play a marginal role, 
enhancing dibaryon binding by merely 15$\pm$5~MeV. Hidden-color considerations 
are of course outside the scope of hadronic models and it is gratifying that 
the results presented here in the hadronic basis are independent of such 
poorly understood configurations.

\acknowledgments 

A.G. acknowledges stimulating correspondence with Mikhail Bashkanov and Heinz 
Clement, and thanks the Organizers of the International Conference HADRON 2013 
for their kind hospitality and for supporting financially his participation in 
this conference.

\end{document}